\documentclass[namedreferences]{solarphysics}
\usepackage[optionalrh]{spr-sola-addons} 
\usepackage{graphicx}        
\usepackage{color}           
\usepackage{url}             

\newcommand{\rmd}{ {\ \mathrm d} }
\renewcommand{\vec}[1]{ {\mathbf #1} }

\newcommand{\an}{     {\it Astron. Nachr.}}

\newcommand{\aap}{    {\it Astron. Astrophys.}}

\newcommand{\apj}{    {\it Astrophys. J.}}
\newcommand{\apjl}{   {\it Astrophys. J. Lett.}}
\newcommand{\apss}{   {\it Astrophys. Space Sci.}}
\newcommand{\gafd}{   {\it Geophys. Astrophys. Fluid Dyn.}}

\newcommand{\mnras}{  {\it Mon. Not. Roy. Astron. Soc.}}

\newcommand{\prl}{    {\it Phys. Rev. Lett.}}
\newcommand{\solphys}{{\it Solar Phys.}}

\newcommand{\ssr}{    {\it Space Sci. Rev.}}
\begin{document}

\begin{article}

\begin{opening}

\title{Solar Dynamo Model with Diamagnetic Pumping and Nonlocal Alpha-Effect}

%
\author{L.L.~Kitchatinov{}$^{1,2}$\sep  S.V.~Olemskoy$^{1}$}

%
\runningauthor{L.L.\,Kitchatinov, S.V.\,Olemskoy}

\runningtitle{Solar Dynamo Model}

%
\institute{$^{1}$ Institute for Solar-Terrestrial Physics, P.O. Box
    291, Irkutsk 664033, Russia, emails: kit@iszf.irk.ru; ovs@iszf.irk.ru \\
    $^{2}$ Pulkovo Astronomical Observatory, St. Petersburg 196140, Russia
 }


\begin{abstract}
A combination of diamagnetic pumping and a nonlocal $\alpha$-effect
of the Babcock-Leighton type in a solar dynamo model helps to
reproduce observations of solar magnetic activity. The period of the
solar cycle can be reproduced without reducing magnetic diffusivity
in the bulk of the convection zone below the standard mixing-length
value of $10^{13}$~cm$^2$s$^{-1}$. The simulated global fields are
antisymmetric about the equator and the toroidal-to-poloidal field
ratio is about a thousand. The time-latitude diagrams of magnetic
fields in the model without meridional flow, however, differ from
observations. Only when the meridional flow is included and the
$\alpha$-effect profile peaking at mid latitudes is applied, can the
observational butterfly diagrams be reproduced.
\end{abstract}

%

\end{opening}

\section{Introduction}
The aim of this paper is to draw attention to the possibility of
resolving several problems of dynamo theory for solar activity by
combining diamagnetic turbulent pumping and a nonlocal
$\alpha$-effect in a solar dynamo model.

In the absence of reliable data on magnetic fields in the deep solar
interior, modeling the solar dynamo remains a controversial issue
(for a recent review, see \opencite{T09}). Consensus has developed
that the dynamo is driven by two basic effects. The $\Omega$-effect
of nonuniform rotation produces a strong toroidal field from a
poloidal one and the $\alpha$-effect of cyclonic motion \cite{P55}
regenerates the poloidal field. Already the first models of an
$\alpha\Omega$-dynamo produced oscillatory solutions resembling the
solar cycle \cite{L69,SK69}. If discrepancies between computed and
observed parameters within two orders of magnitude are tolerable,
these pioneering models are quite satisfactory. A closer agreement
is difficult and may even seem impossible to obtain in view of the
basic physics of the dynamo process and decades-long practice of
dynamo simulations.

Already \inlinecite{K73} noticed that turbulent magnetic diffusivity
should be reduced much below the standard mixing-length value of
$\eta_{_\mathrm{T}} \approx 10^{13}$\,cm$^2$s$^{-1}$ in order to
reproduce the observed period of the solar cycle. The time of
diffusive decay of a magnetic field can be estimated as
$T_\mathrm{d} \approx d^2 \eta^{-1}_{_\mathrm{T}}$, where $d$ is the
depth of the convection zone. A cyclic dynamo has to regenerate
fields in a shorter time in order to overpower the diffusive decay,
so that $P_\mathrm{cyc} < T_\mathrm{d}$ ($P_\mathrm{cyc}$ is the
period of the magnetic cycle). The observed period can be reproduced
with a diffusivity value much smaller than
$10^{13}$\,cm$^2$s$^{-1}$, this case, however, is difficult to
justify. Eddy viscosity or thermal diffusivity below
$10^{13}$\,cm$^2$s$^{-1}$ cannot be a correct parameterization for
solar convection because with such small eddy diffusion the external
layers of the Sun are still unstable \cite{Tea94,KM00}. The
assumption of small magnetic diffusion ne\-eds to explain why the
same turbulent mixing transports momentum and heat much more
efficiently than a magnetic field. Theories of turbulent transport
coefficients do not support the assumption of large magnetic Prandtl
number. Direct numerical simulations of \inlinecite{YBR03} also give
this number of the order of unity.

Observations show a clear predominance of the equator-antisymmetric
(dipolar) component in the global magnetic field of the Sun
\cite{S88}. The symmetric part is relatively small and does not show
an 11-year cycle. The critical dynamo numbers for the excitation of
the global modes of these two types of symmetry are usually very
close together. The threshold dynamo number for the antisymmetric
modes may be relatively small, leading to the preference for this
type of symmetry, but the situation can usually be changed to the
preference for the symmetric modes by small variations of parameters
in a dynamo model. In other words, the equatorial symmetry is
unstable to small changes in the design of dynamo models. Another
problem is related to the ratio of toroidal- to poloidal-field
amplitudes. The ratio is not smaller than a thousand if the poloidal
field is estimated by its polar value and the toroidal field - by
the field strength in sunspots. Solar differential rotation of about
30\% can produce in the 11 years of the solar cycle a toroidal field
that is at most 40 times stronger than the poloidal filed. Strong
radial gradient of rotation in the tachocline does not change this
estimation because the radial field should be as much weaker there
compared to the meridional field as the radial shear is larger than
the latitudinal shear (except for in cases of the relic field
penetrating from the radiation zone, or the field having a structure
with multiple latitudinal belts, both cases present new problems).

Another problem of the so-called catastrophic quenching of the
$\alpha$-effect (Gr\-u\-zinov and Diamond, 1994; \opencite{BS05}) is
not directly related to observations but is \lq internal' for
theory. It can be outlined as follows. The large-scale fields
generated by $\alpha$-effect dynamos are helical. As the magnetic
helicity is conserved, small-scale magnetic fields attain helicity
equal in amount and opposite in sign to that of large-scale fields.
Helical small-scale fields produce their own magnetic
$\alpha$-effect that counteracts the $\alpha$-effect of whatever
origin so that the total $\alpha$-effect strongly diminishes.

In a recent paper \cite{KO11a}, we demonstrated that the
catastrophic quenching is alleviated in a solar dynamo model that
combines a nonlocal $\alpha$-effect and diamagnetic pumping of
large-scale fields by turbulent convection. In this new publication
it is shown that with this model other above-mentioned problems also
fade. The diamagnetic pumping, which will be discussed in the next
section, concentrates the magnetic field at the bottom of the
convection zone. The turbulent diffusion in this region is small. As
a result, the solar cycle period can be reproduced. The poloidal
field in the near-base region is much stronger than on the surface
and sufficiently strong for the differential rotation to generate
kilogauss toroidal fields over a solar cycle. The critical dynamo
number for dipolar dynamo modes is considerably smaller compared to
quadrupolar modes, so that the model always results in a global
field of dipolar parity. However, the model still has difficulties
in reproducing time-latitude diagrams of solar magnetism. We include
the meridional flow and vary the latitudinal profile of the
$\alpha$-effect to reproduce observational butterfly diagrams.
\section{Diamagnetic Pumping}\label{diamagnetism}
Diamagnetic pumping of large-scale fields was predicted
theoretically long time ago \cite{Z57,R68} but is scarcely known in
the dynamo community and is usually ignored in dynamo models (see,
however, R\"udiger and Bran\-denburg, 1995; \opencite{KKT06};
Guerrero and de Gouveia Dal Pino, 2008).

The nature of diamagnetic pumping is well illustrated by the case of
inhomogeneous 2D turbulence with the fluctuating velocity $\vec{u}'$
uniform along a direction defined, say, by the unit vector $\vec{e}$
and inhomogeneous in the direction perpendicular to $\vec{e}$:
${\bf\nabla}\langle {u'}^2\rangle \neq 0$, $
(\vec{e}\cdot{\bf\nabla})\vec{u'} = 0$. For this case,
\inlinecite{Z57} found that the mean electromotive force, ${\bf\cal
E} = \langle \vec{u}'\times\mathrm{\bf B}'\rangle$ ($\mathrm{\bf
B}'$ is a fluctuating magnetic field), can be written as
\begin{equation}
    \bf{\cal E} = -{\bf\nabla}\times\left( \eta_{_\mathrm{T}}\vec{B}\right),
    \label{1}
\end{equation}
if the large-scale field $\vec{B}$ is also perpendicular to
$\mathrm{\bf e}$; $\eta_{_\mathrm{T}}$ is the turbulent magnetic
diffusivity. This means that the large-scale field is transported
with the effective velocity
\begin{equation}
    \mathrm{\bf U}_\mathrm{dia} = -{\bf\nabla}\eta_{_\mathrm{T}} .
    \label{2}
\end{equation}
The diamagnetic pumping expels the field from the regions of
relatively high turbulence intensity. However, for the case of the
mean field parallel to the direction $\mathrm{\bf e}$, the mean
electromotive force reads
\begin{equation}
    {\bf\cal E} = -\eta_{_\mathrm{T}}{\bf\nabla}\times\vec{B}
    \label{3}
\end{equation}
and there is no diamagnetic pumping. The magnetic field parallel to
the $\vec{e}$-direction, $\vec{e}\|\vec{B}$, behaves like a scalar
field. The diamagnetic pumping, therefore, is related to the
vectorial nature of magnetic fields, and there is no counterpart of
diamagnetic pumping for scalar fields.

 \begin{figure}[htb]
 \begin{center}
 \includegraphics[width=0.55\textwidth]{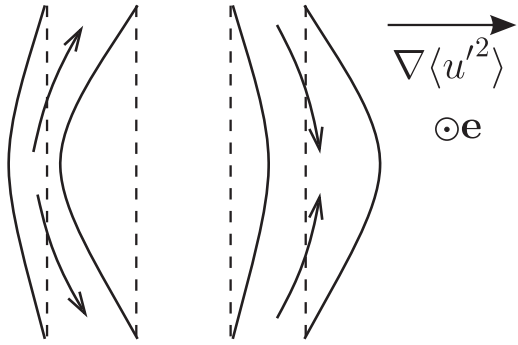}
 \end{center}
 \caption[]{
Pictorial explanation of diamagnetic pumping (see text). Turbulent
intensity increases from left to right. Undisturbed magnetic lines
are shown by dashed lines. When a flux tube is displaced by motion,
its right boundary is moving faster on average. Displacements to the
left compress the flux tube and the field strength in the displaced
region increases. Material is evacuated from the displaced region as
shown by arrows. Displacements to the right result on average in a
decrease in field strength in the displaced region.
 }
 \label{f1}
 \end{figure}

The pumping effect can be interpreted as follows (Figure~\ref{f1}).
Imagine that there is a background field perpendicular to both the
direction $\mathrm{\bf e}$ and the gradient of turbulence intensity.
If the turbulent motion displaces a flux tube of such a field as the
one on the left in Figure~\ref{f1}, {\it i.e.}, in the direction of
decreasing turbulence intensity, the left boundary of the tube is on
average moving slower than the right boundary. The tube is
compressed, and the field strength in the displaced region
increases. Similarly, displacement to the right decreases the field
strength in the displaced region. Turbulent mixing produces on
average the field transport in the direction of decreasing turbulent
intensity, {\it i.e.}, turbulent conducting fluids behave
diamagnetic.

In the 3D case, the expression for the effective velocity of
diamagnetic transport changes to \cite{KR80}
\begin{equation}
    \vec{U}_\mathrm{dia} = -\frac{1}{2}{\bf\nabla}\eta_{_\mathrm{T}} .
    \label{4}
\end{equation}
The diamagnetic pumping can be very efficient near the bottom of the
convection zone where the intensity of turbulent convection changes
sharply with depth \cite{KR08}.

Doubts have been expressed concerning the efficiency of the pumping
in a nonlinear regime \cite{VK83}. However, turbulent diamagnetism
was recently detected in laboratory experiments with liquid sodium
\cite{Sea07}. Direct numerical simulations also show downward
pumping of large-scale fields near the base of convectively unstable
layers (Tobias {\it et al.}, 1998; \opencite{Tea98,DN01,Oea02};
Zigler and R\"udiger, 2003). Whether the simulations indeed show the
turbulent diamagnetism is not perfectly clear. Their results were
interpreted in terms of the Drobyshevski and Yuferev (1974) effect
of topological pumping (\opencite{Tea01}; Dorch and Nordlund, 2001).
Anyway, the diamagnetic and topological pumpings are of the same
sense (downward) near the base of the convection zone. We apply the
Equation\,(\ref{4}) for the effective transport velocity in our
dynamo model.

It may be noted that turbulent diffusion and pumping are both
reduced by rotation and/or a magnetic field. \inlinecite{Tea01} and
\inlinecite{ZR03} observed the rotational quenching effect in
numerical simulations. An analytical theory of turbulent transport
predicts that though the turbulent diffusion and pumping are
modified, the relation (\ref{4}) between their isotropic parts
survives under the influence of rotation or a magnetic field
\cite{K88}.
\section{The Model}
\subsection{Dynamo Equations}
Our dynamo model is based on the mean-field induction equation
\begin{equation}
    \frac{\partial\vec{B}}{\partial t} =
    {\bf\nabla}\times\left( \vec{V} \times \vec{B}
    + {\bf\cal E}\right) .
    \label{5}
\end{equation}
The large-scale flow is a superposition of rotation and meridional
circulation,
\begin{equation}
    \vec{V} = \vec{e}_\phi r\sin\theta\Omega f(r,\theta) +
    \frac{1}{\rho}{\bf\nabla}\times\left( \vec{e}_\phi
    \frac{\psi}{r\sin\theta}\right) ,
    \label{6}
\end{equation}
where the usual spherical coordinates are used, $\Omega$ is the
characteristic value of angular velocity, $f$ is the normalized
frequency of differential rotation, $\psi$ is the meridional flow
stream function, $\rho$ is density, and $\vec{e}_\phi$ is the
azimuthal unit vector.

The mean electromotive force is written as
\begin{equation}
   {\bf\cal E} = -\sqrt{\eta_{_\mathrm{T}}}\ {\bf\nabla}\times\left(
   \sqrt{\eta_{_\mathrm{T}}} \vec{B}\right) + {\bf\cal A} ,
   \label{7}
\end{equation}
where the first term on the right stands for both turbulent
diffusion and the diamagnetic pumping of Equation\,(\ref{4}). The
second term accounts for the $\alpha$-effect in its nonlocal
formulation \cite{BRS08}
\begin{equation}
  {\bf\cal A} = \int{\alpha}(\mathrm{\bf r},\mathrm{\bf
  r}')\vec{B}(\mathrm{\bf r}')\ \mathrm{d}^3r' .
  \label{8}
\end{equation}
The local $\alpha$-effect is subject to the catastrophic quenching
mechanism (Bran\-denburg and Subramanian, 2005) while the nonlocal
$\alpha$-effect is not (Kitchatinov and Olemskoy, 2011a), and for
this reason it may be dominating in the dynamo process. Another
possibility to avoid the catastrophic quenching is to invoke
sufficiently efficient fluxes of magnetic helicity (Guerrero,
Chatterjee, and Brandenburg, 2010).

Similar to the mean flow of Equation\,(\ref{6}), the magnetic field
can be written as a superposition of its toroidal and poloidal
parts,
\begin{equation}
    \vec{B} = \vec{e}_\phi B + {\bf\nabla}\times
    \left( \vec{e}_\phi\frac{A}{r\sin\theta}\right) ,
    \label{9}
\end{equation}
where $A$ is the poloidal field potential.

Here we introduce normalized variables. The time is measured in
units of $R_\odot^2/\eta_0$; $\eta_0$ is the characteristic value of
the eddy diffusivity. The magnetic field is normalized to the field
strength $B_0$ for which nonlinear effects become essential, and the
$\alpha$-parameter is normalized to its characteristic value
$\alpha_0$. The poloidal field potential is measured in units of
$\alpha_0B_0R^3_\odot/\eta_0$. The density is normalized to its
surface value $\rho_0$, and the stream function of meridional flow
is measured in units of $\rho_0 R_\odot^2V_0$; $V_0$ is the
amplitude of the surface meridional flow. From now on, the same
notations are kept for the normalized variables as used before for
their unnormalized counterparts, except for the fractional radius $x
= r/R_\odot$ and normalized diffusivity $\eta =
\eta_{_\mathrm{T}}/\eta_0$. The normalized equation for the toroidal
field reads
 \begin{eqnarray}
    \frac{\partial B}{\partial t} &=&
    \frac{\eta}{x^2}\frac{\partial}{\partial\theta}\left(
    \frac{1}{\sin\theta}\frac{\partial(\sin\theta
    B)}{\partial\theta}\right)
    + \frac{1}{x}\frac{\partial}{\partial
    x}\left(\sqrt{\eta}\ \frac{\partial(\sqrt{\eta}\ xB)}
    {\partial x}\right) +
    \nonumber \\
    &+& \frac{R_\mathrm{m}}{x}\frac{\partial}{\partial\theta}
    \left(\frac{B}{\rho x\sin\theta}
    \frac{\partial\psi}{\partial x}\right)
    - \frac{R_\mathrm{m}}{x}\frac{\partial}{\partial x}
    \left(\frac{B}{\rho x\sin\theta}\frac{\partial\psi}{\partial\theta}
    \right) +
    \nonumber \\
    &+& \frac{\cal D}{x}
    \left(\frac{\partial f}{\partial x}\frac{\partial
    A}{\partial\theta} - \frac{\partial f}{\partial\theta}
    \frac{\partial A}{\partial x}\right) ,
 \label{10}
 \end{eqnarray}
where
 \begin{equation}
    {\cal D} = \frac{\alpha_0 \Omega R_\odot^3}{\eta_0^2}\
 \label{11}
 \end{equation}
is the dynamo number and
\begin{equation}
    R_\mathrm{m} = \frac{V_0 R_\odot}{\eta_0}
    \label{12}
\end{equation}
is the magnetic Reynolds number for the meridional flow. The
$\alpha\Omega$-approximation is applied to neglect the alpha-effect
in the toroidal field Equation (\ref{10}). This equation describes
the toroidal-field production by differential rotation, its
advection by the meridional flow, diamagnetic pumping, and turbulent
diffusion.

The poloidal-field equation with nonlocal $\alpha$-effect is written
as
 \begin{eqnarray}
    \frac{\partial A}{\partial t} &=&
    \frac{\eta}{x^2}\sin\theta\frac{\partial}{\partial\theta}
    \left(\frac{1}{\sin\theta}\frac{\partial
    A}{\partial\theta}\right) + \sqrt{\eta}\frac{\partial}{\partial
    x} \left(\sqrt{\eta}\frac{\partial A}{\partial x}\right) +
    \nonumber \\
    &+& \frac{R_\mathrm{m}}{\rho x^2 \sin\theta}
    \left(\frac{\partial\psi}{\partial x}
    \frac{\partial A}{\partial\theta} -
    \frac{\partial\psi}{\partial\theta}
    \frac{\partial A}{\partial x}\right) +
    \nonumber \\
    &+& x \sin\theta
    \cos\theta \int\limits_{x_\mathrm{i}}^x
    \hat\alpha (x,x') B(x',\theta)\ \mathrm{d} x'  ,
 \label{13}
 \end{eqnarray}
where $x_\mathrm{i}$ is the radius of the inner boundary. The
integration in this equation is only in the radius with the upper
limit $x$. This qualitatively reflects the fact that the nonlocal
$\alpha$-effect at some point $x$ is contributed by the buoyant
magnetic loops rising from deeper layers ($x' < x$) and that buoyant
velocities are almost vertical. Our dynamo Equations (\ref{10}) and
(\ref{13}) are very similar to those in our previous publication
\cite{KO11a}, but now we include meridional flow and neglect the
(local) magnetic $\alpha$-effect which was shown to be insignificant
when nonlocal $\alpha$-effect is allowed for.

Our boundary conditions assume a perfect conductor beneath the inner
bo\-un\-da\-ry of radius $x_\mathrm{i}$ and pseudo-vacuum conditions
on the top,
\begin{eqnarray}
    \frac{\partial\left(\sqrt{\eta} x B\right)}{\partial x} &=& 0,\ \
    A = 0\ \ \ \ \mathrm{for}\ \  x = x_\mathrm{i} ,
    \nonumber \\
    \frac{\partial A}{\partial x} &=& 0,\ \  B = 0,\ \ \ \
    \mathrm{for}\ \  x = 1 .
    \label{14}
\end{eqnarray}
The bottom boundary in our model is at $x_\mathrm{i} = 0.7$.

The initial-value problem for the dynamo Equations (\ref{10}) and
(\ref{13}) was solved numerically by the grid-point method and
explicit time-stepping. The diamagnetic pumping leads to a high
concentration of the magnetic field near the bottom. To resolve fine
structures near the bottom, a nonuniform grid was applied over the
radius with the grid spacing $\Delta x \sim \eta^{1/2}$. The grid
over the latitude was uniform.

The equatorial symmetry was usually not prescribed. The field was
evolved in time starting from a mixed-parity initial field and the
solution relaxed eventually to a certain equatorial symmetry. In
order to determine the critical dynamo numbers for the excitation of
the dipolar ($B(\theta ) = -B(\pi-\theta)$) and quadrupolar
($B(\theta ) = B(\pi-\theta)$) dynamo modes, additional boundary
conditions selecting the field mode of certain equatorial symmetry
were imposed on the equator.
\subsection{Model Design}
\subsubsection{Differential Rotation}
\begin{figure}[htb]
 \begin{center}
 \includegraphics[width=0.5\textwidth]{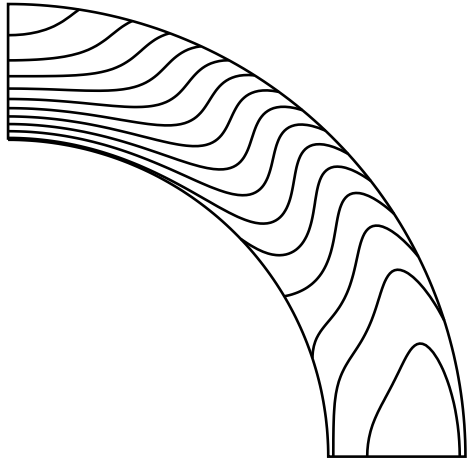}
 \end{center}
 \caption[]{
Angular velocity contours for the differential rotation used in our
dynamo model.
 }
 \label{f2}
 \end{figure}

For the differential rotation, we use the approximation of
helioseismological data suggested by \inlinecite{BKS00}
 \begin{equation}
    f(x,\theta) = \frac{1}{461}\sum\limits_{m=0}^{2}
    \cos\left( 2m\left(\frac{\pi}{2}
    - \theta\right)\right) \sum\limits_{n=0}^{4} C_{nm}x^n .
    \label{15}
 \end{equation}
The coefficients $C_{nm}$ of this equation are given in Table~1 of
Belvedere, Kuzan\-yan, and Sokoloff (2000). Figure~\ref{f2} shows
the angular velocity contours.
\subsubsection{Profiles of the $\alpha$-Effect and Diffusivity}
The kernel function of the nonlocal $\alpha$-effect in the poloidal
field Equation (\ref{13}) was prescribed as follows,
\begin{eqnarray}
    \hat\alpha (x,x') &=& \frac{\phi_\mathrm{b}(x')\phi_\alpha (x)} {1 + B^2(x',\theta)}
    ,
    \nonumber \\
    \phi_\mathrm{b}(x') &=& \frac{1}{2}\left( 1 -
    \mathrm{erf}\left( (x' - x_\mathrm{b})/h_\mathrm{b}\right)\right) ,
    \nonumber \\
    \phi_\alpha (x) &=& \frac{1}{2}\left( 1 +
    \mathrm{erf}\left( (x - x_\alpha)/h_\alpha\right)\right),
    \label{16}
\end{eqnarray}
where $\mathrm{erf}$ is the error function and $B^2$ in the
denominator of the first equation accounts for the usual algebraic
quenching of the $\alpha$-effect. We always use $x_\mathrm{b} =
x_\mathrm{i} + 2.5h_\mathrm{b}$ and $x_\alpha = 1 - 2.5h_\alpha$ to
ensure smoothness of the kernel functions in the simulation domain.
The parameter $h_\mathrm{b}$-parameter represents the thickness of
the near-bottom region of toroidal magnetic fields producing the
$\alpha$-effect. The parameter $h_\alpha$ represents the thickness
of the near-surface layer where this effect is produced. The
$\alpha$-effect with the kernel function of Equation (\ref{16}) is
very close to the nonlocal model of \inlinecite{BK07}. It is also
similar to the Babcock-Leighton mechanism for the poloidal field
production used in the dynamo models of \inlinecite{D95} and
\inlinecite{DC99}.

The turbulent diffusivity in the bulk of the convection zone varies
slightly with depth. Near the base of the convection zone, it drops
sharply with increasing depth. The diffusivity profile of our model
reads
\begin{equation}
    \eta (x) = \eta_\mathrm{in} + \frac{1}{2}(1 - \eta_\mathrm{in})
    \left( 1 + \mathrm{erf}\left(\frac{x -
    x_\eta}{h_\eta}\right)\right) ,
    \label{17}
\end{equation}
where $\eta_\mathrm{in}$ is the ratio of diffusivity near the inner
boundary to its value in the bulk of the convection zone.
Computations were performed for the smallest value of
$\eta_\mathrm{in} = 10^{-4}$ we were able to apply.

 \begin{figure}[htb]
 \begin{center}
 \includegraphics[width=0.6\textwidth,height=0.5\textwidth]{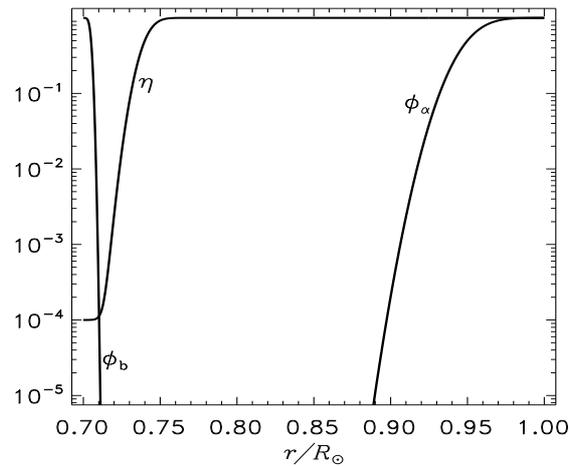}
 \end{center}
 \caption[]{
    Profiles of the normalized diffusivity and the kernel functions
    of Equation (\ref{16}) of the nonlocal $\alpha$-effect.
 }
 \label{f3}
 \end{figure}

The major part of our computations was performed with the following
values of parameters: $x_\eta = 0.74$, $h_\eta = 0.01$, $h_\alpha =
0.02$, and $h_\mathrm{b} = 0.002$. The profiles of diffusivity and
kernel functions of the $\alpha$-effect of Equation (\ref{16}) for
this set of parameters are shown in Figure~\ref{f3}. Dependence of
the results on the model parameters $h_\mathrm{b}$ and $h_\alpha$
will be discussed at the beginning of Section~\ref{RD} to explain
this choice.
\subsubsection{Meridional Flow}
A poleward meridional flow of the order of 10~m\,s$^{-1}$ is
observed on the solar surface \cite{KHH93}. Helioseismology confirms
that the flow persists up to a depth of about 12\,Mm \cite{ZK04}.
Theoretical modeling remains the only source of knowledge about the
flow in deeper regions. Recent simulations predict that one cell of
meridional circulation occupies the entire thickness of the
convection zone and the return flow at the bottom is not small
compared to the surface \cite{KO11b}. Beneath the convection zone,
the flow is small (\opencite{GM04}; Kitchatinov and R\"udiger,
2006). All these findings are qualitatively reflected by the
following representation for the stream function of the meridional
flow
\begin{eqnarray}
    \psi &=&- \cos\theta\ \sin^2\theta\ \phi(x) ,
    \nonumber  \\
    \phi(x) &=& \left\{\begin{array}{ll}
            \frac{1}{1 - x_\mathrm{s}} \int\limits_x^1 \rho
            (x')\eta^p(x')(x'-x_\mathrm{s}) \rmd x' & \mbox{for $x
            \geq x_\mathrm{s}$} \\
            C \int\limits_{x_\mathrm{i}}^x \rho(x') \eta^p(x')
            (x_\mathrm{s} - x') \rmd x' & \mbox{for $x
            \leq x_\mathrm{s} .$}
            \end{array}
            \right.
            \label{18}
\end{eqnarray}
In this equation, $x_\mathrm{s}$ is the radius of the stagnation
point where the meridional velocity changes sign, and $C$ is a
parameter whose value is adjusted to ensure continuity of the stream
function at the stagnation point.

 \begin{figure}[htb]
 \begin{center}
 \includegraphics[width=0.6\textwidth,height=0.41\textwidth]{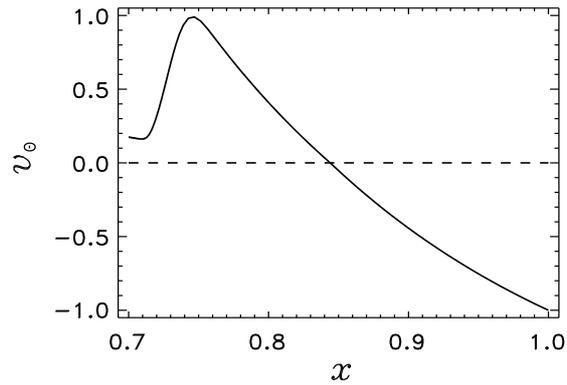}
 \end{center}
 \caption[]{
Radial profile of meridional velocity in units of $V_0/2$ at a
latitude of 45$^\circ$.
 }
 \label{f4}
 \end{figure}

The stratification of the convection zone is almost adiabatic. An
adiabatic profile for ideal gas was used for the normalized density,
\begin{equation}
    \rho(x) = \left( 1 + C_\rho\left(\frac{1}{x}
    -1\right)\right)^{3/2} ,
    \label{19}
\end{equation}
where $C_\rho = 10^3$.

The profile of the normalized meridional velocity of our model is
shown in Figure~\ref{f4}. The radius of the stagnation point,
$x_\mathrm{s} \simeq 0.84$, was adjusted so that the amplitudes of
the flow above and below this point are equal. We set the parameter
$p = 0.25$ in Equation~(\ref{18}), so that the bottom flow is about
ten times slower than the maximum velocity of the deep equator-ward
flow.
\section{Results and Discussion}\label{RD}
\subsection{Models without Meridional Flow}
We will discuss the models without meridional flow ($R_\mathrm{m}$ =
0) first. Then, we will include the flow to see what effect it
produces.

 \begin{figure}[htb]
 \centerline{
 \includegraphics[width=0.4\textwidth,height=0.4\textwidth]{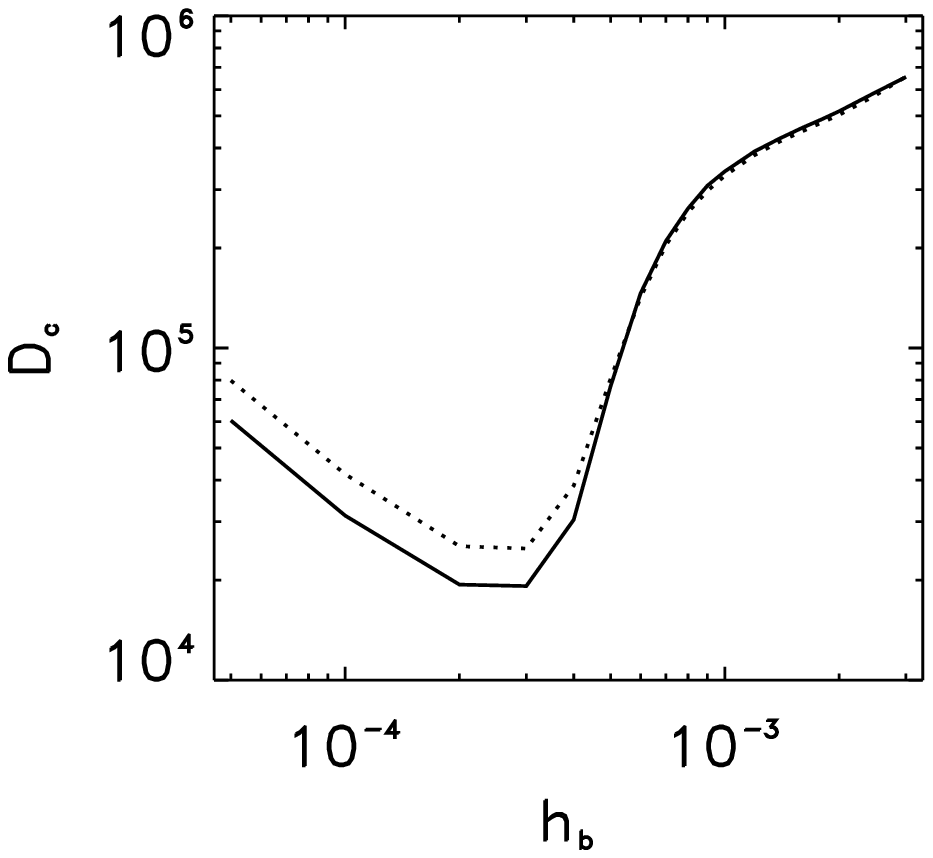}
 \hspace{0.05\textwidth}
 \includegraphics[width=0.4\textwidth,height=0.4\textwidth]{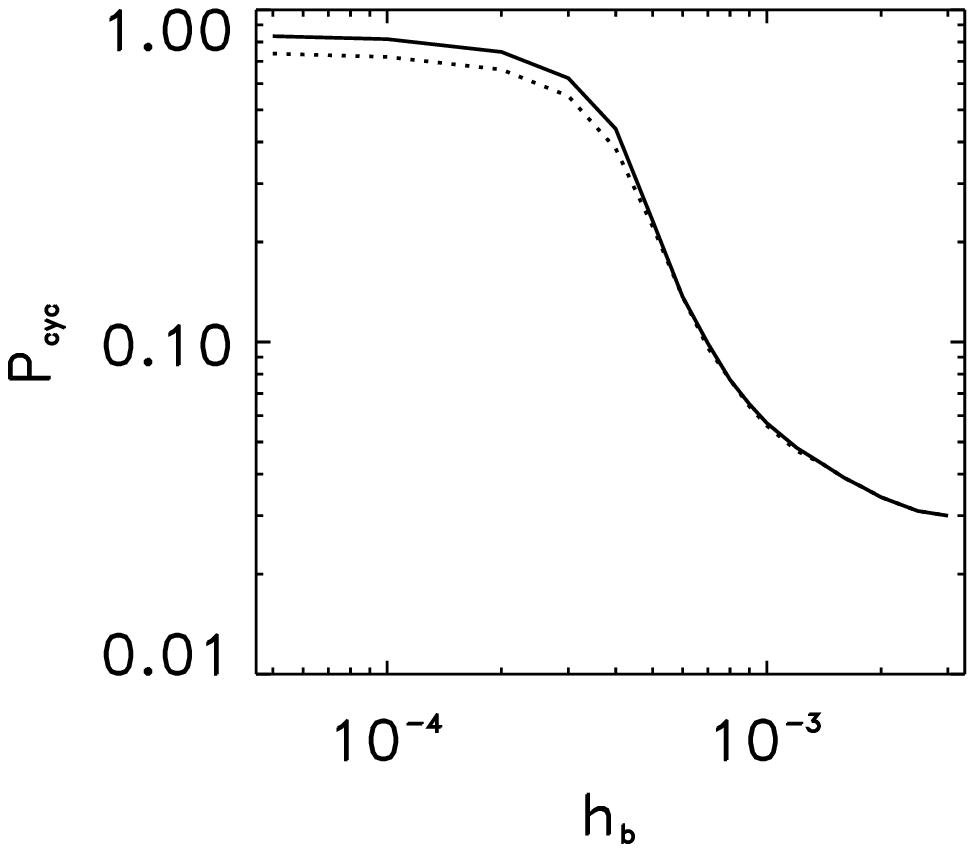}
  }
 \vspace{-0.31\textwidth}   
     \centerline{
      \hspace{0.22\textwidth}   \color{black}{(a)}
      \hspace{0.42\textwidth}   \color{black}{(b)}
         \hfill}
     \vspace{0.3\textwidth}    
 \caption[]{
Dependence of the critical dynamo number (a) and the magnetic cycle
period (b) on the thickness $h_\mathrm{b}$ of the near-bottom layer
producing the $\alpha$-effect. Full and dotted lines show the
results for dipolar and quadrupolar dynamo modes, respectively.
 }
 \label{f5}
 \end{figure}

Figure~\ref{f5} shows the dependencies of the critical dynamo number
and magnetic cycle period on the thickness $h_\mathrm{b}$ of the
near-bottom layer producing the $\alpha$-effect (for constant
$h_\mathrm{t} = 0.02$). The cycle period $P_\mathrm{cyc}$ is defined
as half the complete period of magnetic oscillations in order to
compare with the 11-year solar cycle. The dipolar parity and the
observed period of magnetic oscillations can be reproduced only with
a sufficiently thin bottom layer, $h_\mathrm{b} < 4\times 10^{-3}$.
Only in this case the equator-antisymmetric modes of the magnetic
field are preferred and the cycle period is sufficiently long. For
the value $\eta_0 = 10^{13}$cm$^2$s$^{-1}$ of magnetic diffusivity,
the diffusion time $R^2_\odot/\eta_0 \approx 15$~years and the cycle
period in physical units, $P_\mathrm{cyc}R^2_\odot/\eta_0$, is close
to 11 years for small $h_\mathrm{b}$. Also for a thin bottom layer
only, the critical dynamo number for the excitation of magnetic
fields of dipolar parity is about 30\% smaller compared to
quadrupolar parity modes so that the model produces
equator-antisymmetric global fields similar to the Sun. A strong
decrease of magnetic diffusivity towards the bottom boundary is also
important for preference of dipolar parity \cite{CNC04,HY10}. We
conclude that our model can reproduce a solar cycle only if the
$\alpha$-effect is produced by sufficiently deep-seated toroidal
fields.

 \begin{figure}[htb]
 \centerline{
 \includegraphics[width=0.4\textwidth,height=0.4\textwidth]{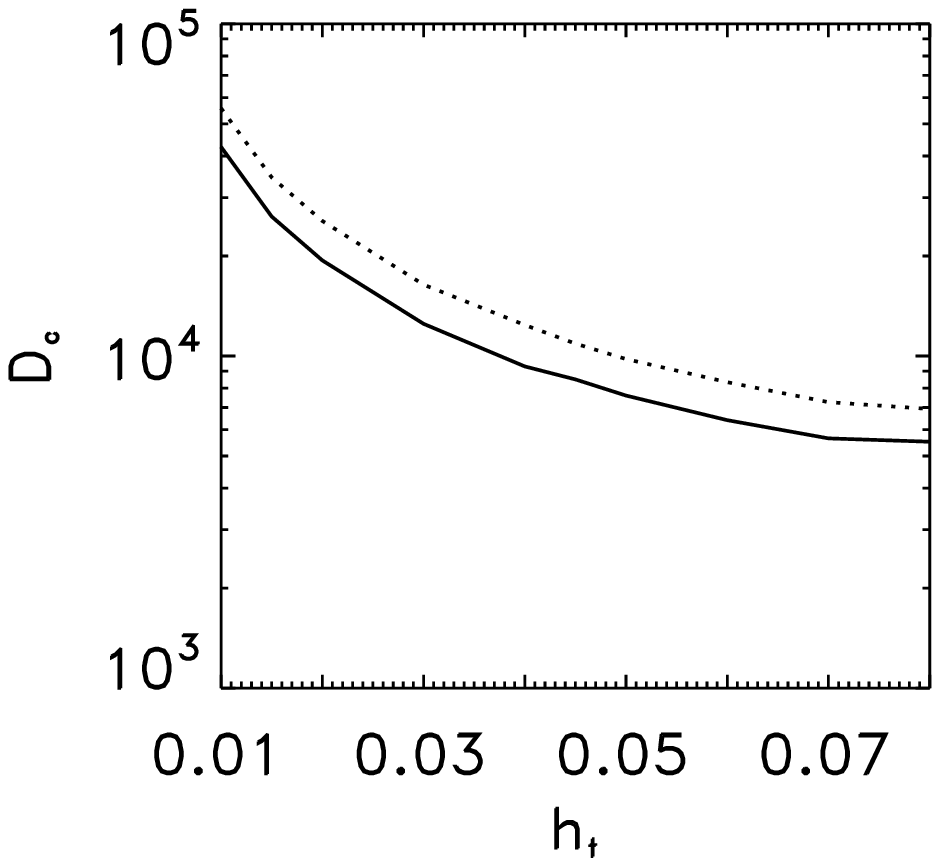}
 \hspace{0.05\textwidth}
 \includegraphics[width=0.4\textwidth,height=0.4\textwidth]{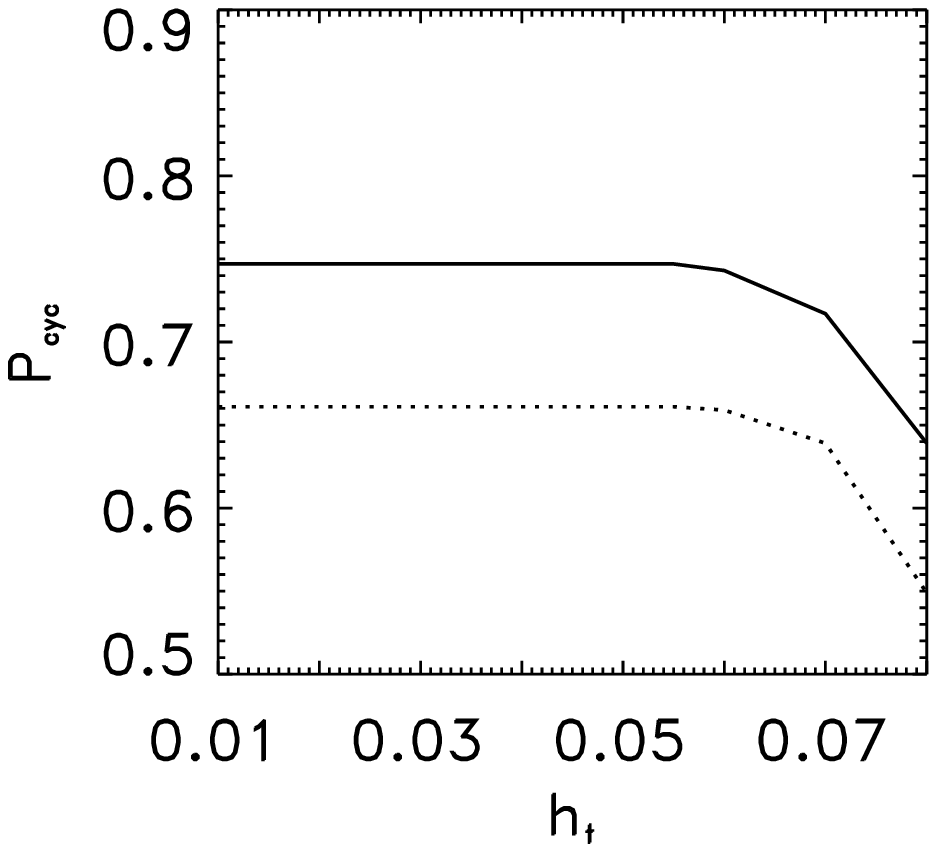}
  }
 \vspace{-0.34\textwidth}   
     \centerline{
      \hspace{0.22\textwidth}   \color{black}{(a)}
      \hspace{0.42\textwidth}   \color{black}{(b)}
         \hfill}
     \vspace{0.33\textwidth}    
 \caption[]{
Dependence of the critical dynamo number (a) and the magnetic cycle
period (b) on the thickness $h_\mathrm{t}$ of the near-top layer
where the $\alpha$-effect is produced. Full and dotted lines show
the results for dipolar and quadrupolar modes of magnetic field,
respectively. $h_\mathrm{b} = 2\times 10^{-3}$.
 }
 \label{f6}
 \end{figure}

The dynamo numbers shown in Figure~\ref{f5} have a minimum at about
$h_\mathrm{b} = 2\times 10^{-3}$, and all the results to follow were
produced with this value. For smaller $h_\mathrm{b}$, the critical
dynamo numbers are larger because of the decrease in the toroidal
magnetic flux producing the $\alpha$-effect. For larger
$h_\mathrm{b}$, the bottom layer is thicker than the vertical scale
of the toroidal field and the fields of opposite signs partly cancel
each other's contribution to the $\alpha$-effect. The dependence on
the thickness $h_\mathrm{t}$ of the top layer where the
$\alpha$-effect is produced is simpler. The cycle period shown in
Figure~\ref{f6} does not depend on $h_\mathrm{t}$ until the
thickness becomes so large that the top and bottom layers overlap.
There is also a smooth decrease in the critical dynamo numbers with
increasing $h_\mathrm{t}$. We fix $h_\mathrm{t} = 0.02$ for further
computations.

 \begin{figure}[htb]
 \centerline{
 \includegraphics[width=0.9\textwidth,height=0.7\textwidth]{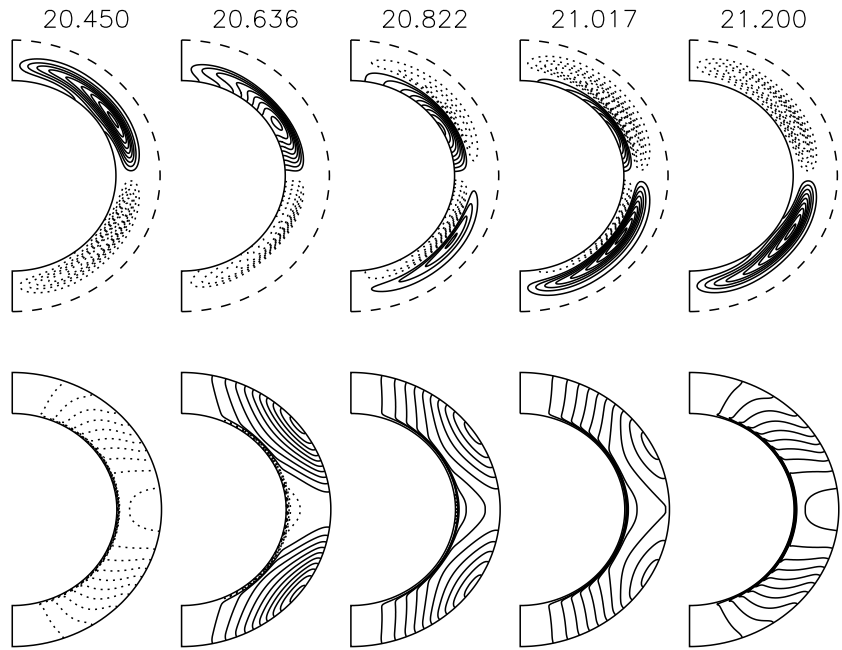}
 }
 \caption[]{
    Contours of toroidal-field (top row) and poloidal field lines
    (bottom row) for several instances of a magnetic cycle. The time of the
    run in units of $R^2_\odot/\eta_0$ is shown at the top. Full
    (dotted) lines show positive (negative) levels and clockwise
    (anticlockwise) circulation. The pictures of the upper row are
    re-scaled so that the upper (dashed) boundary shows the radius of
    $r = 0.74R_\odot$, below which the toroidal fields are localized.
    The dynamo number $D = 2.2\times 10^4$ is slightly above the critical value of
    $1.9\times 10^4$.
 }
 \label{f7}
 \end{figure}

Figure~\ref{f7} shows the magnetic field patterns in the meridional
cross-section for several instances of a cycle. The magnetic field
is highly concentrated at the bottom. The concentration is provided
by the diamagnetic pumping. The pumping, however, does not affect
the (radial) component of the field parallel to the pumping
direction and the poloidal field lines can come to the surface. The
toroidal field is confined below the radius of $x_\eta$ where the
magnetic diffusivity drops sharply with depth. The poloidal field
near the bottom is also much stronger than at the top. As a result,
the toroidal field amplitude in our model is about a thousand times
stronger than the amplitude of the surface polar field. If the
poloidal field at the bottom were of the same order as at the top,
the differential rotation would not be able to produce a strong
toroidal field over a cycle. Note that the deep toroidal field of
several kilogauss in our model is the {\em mean} field. The total
field can include strong fluctuations on the background of the mean
field.

 \begin{figure}[htb]
 \begin{center}
 \includegraphics[width=0.8\textwidth,height=0.55\textwidth]{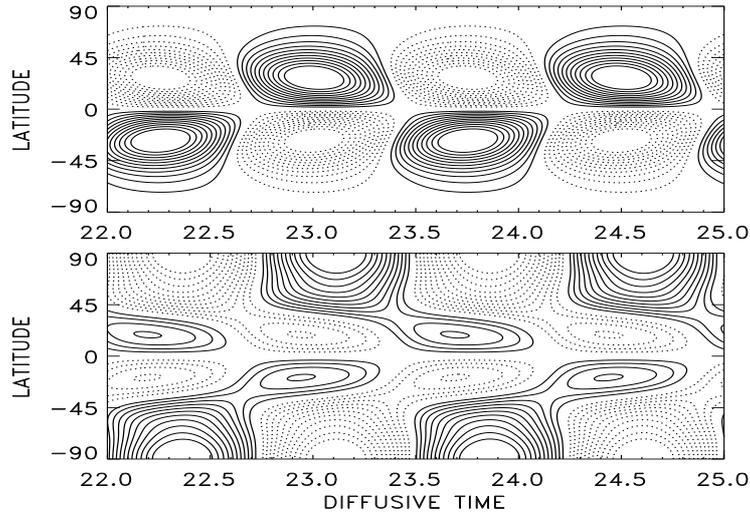}
 \end{center}
 \caption[]{
    Butterfly diagram of the depth-integrated toroidal field
    $\cal B$ of Equation (\ref{20}) (top panel) and the surface radial field
    (bottom) for the model without meridional flow. Time is shown in
    units of $R^2_\odot /\eta_0$.
    $D = 2.2\times 10^4$.
 }
 \label{f8}
 \end{figure}

Figure~\ref{f8} shows butterfly diagrams for the surface radial
field and deep toroidal field. The toroidal field diagram shows the
contours of the quantity
\begin{equation}
    {\cal B} = \sin\theta \int\limits_{x_\mathrm{i}}^1
    \phi_\mathrm{b}(x) B(x)\ \mathrm{d}x ,
    \label{20}
\end{equation}
to which the surface $\alpha$-effect of our model is proportional.
The factor $\sin\theta$ in Equation (\ref{20}) accounts for the
dependence of the length of toroidal flux tubes on latitude (it is
supposed that the probability of sunspot production is proportional
to the length of the tube).

 \begin{figure}[htb]
 \begin{center}
 \includegraphics[width=0.8\textwidth,height=0.55\textwidth]{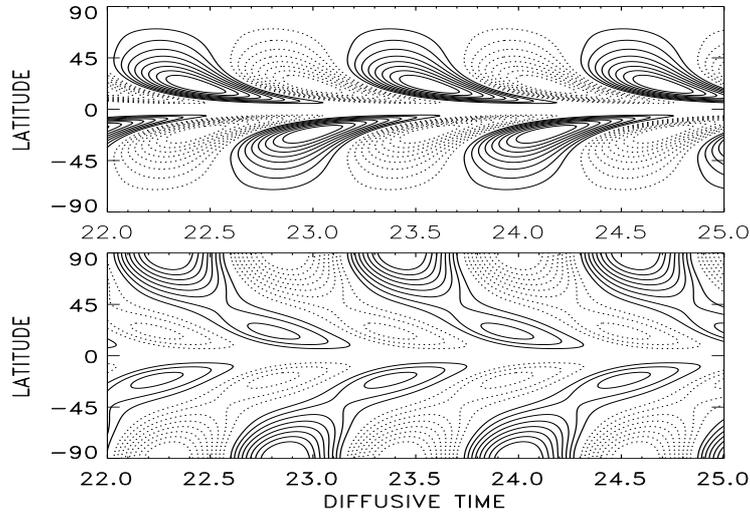}
 \end{center}
 \caption[]{
    Time-latitude diagrams for the model with meridional flow. A standard
    latitudinal profile of the $\alpha$-effect, $\alpha\sim\cos\theta$, was used. The dynamo
    number $D = 2.6\times 10^4$ is slightly above the critical value of
    $D_\mathrm{c} = 2.36\times 10^4$. The top and bottom panels show
    the deep toroidal field and the surface radial field respectively.
 }
 \label{f9}
 \end{figure}

The radial field diagram of Figure~\ref{f8} is similar to
observational diagrams of \inlinecite{S88} and \inlinecite{Oea06}.
The toroidal field diagram, however, has too broad \lq wings' and
shows too slow an equatorial drift compared to the butterfly diagram
of sunspots. There is a possibility to improve the agreement with
allowance for meridional flow.
\subsection{Models with Meridional Flow}
All computations with meridional flow were performed for the
magnetic Reynolds number $R_\mathrm{m} = 10$ that is a plausible
solar value for the magnetic diffusivity $\eta_0 =
10^{13}$~cm$^2$\,s$^{-1}$. The \lq local' Reynolds number in the
near-bottom region of low diffusivity is, of course, much larger. It
may be expected, therefore, that the meridional flow influences
primarily the field migration near the bottom. Indeed, the toroidal
field diagram of Figure~\ref{f9} shows much faster equatorial drift
compared to Figure~\ref{f8}.

 \begin{figure}[hbt]
 \begin{center}
 \includegraphics[width=0.8\textwidth,height=0.55\textwidth]{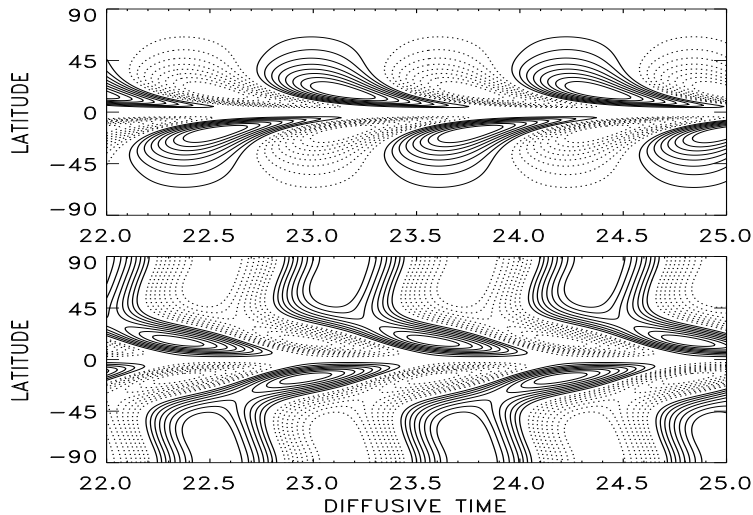}
 \end{center}
 \caption[]{
    Time-latitude diagrams for the model with meridional flow and
    $\sin^2\theta\cos\theta$-profile of the $\alpha$-effect. Computations
    were performed for the dynamo number $D = 4.2\times 10^4$, slightly above
    the critical value of $D_\mathrm{c} = 3.96\times 10^4$. The top and bottom panels
    show the deep toroidal field and the surface radial field respectively.
 }
 \label{f10}
 \end{figure}

Now, however, the poloidal field diagram becomes unsatisfactory. It
shows equatorial drift also at high latitudes where the poleward
migration is actually observed \cite{S88}. The magnetic diffusion
near the surface is high and the meridional flow does not influence
the surface fields directly. However, the surface poloidal fields
are related to the deep toroidal fields by the $\alpha$-effect and
follow the equatorial migration of the deep fields.

The polar drift of the high-latitude poloidal field can be
re-established by a change in the latitudinal profile of the
$\alpha$-effect. Observations indicate that the current helicity of
solar active regions (the $\alpha$-effect proxy) does not increase
steadily with latitude but has a maximum at mid latitudes (Sokoloff
{\it et al.}, 2008). Theoretical tilt-angles of bipolar magnetic
regions show also humps in the latitudinal profiles \cite{DC93}.
Figure~\ref{f10} shows the model results for the
$\sin^2\theta\cos\theta$-profile of the $\alpha$-effect, {\it i.e.},
$\sin\theta$ in the last term of Equation~(\ref{13}) was changed to
$\sin^3\theta$. With this profile, the poloidal field diagram is in
closer agreement with observations.

Simulations that take into account the meridional flow also result
in magnetic fields of dipolar parity. The critical dynamo numbers
for the excitation of dipolar and quadrupolar dynamo modes in the
model with the $\alpha$-effect peaking at mid latitudes ($\alpha
\sim \sin^2\theta\cos\theta$) are $D_\mathrm{c}^\mathrm{d} =
3.96\times 10^4$ and $D_\mathrm{c}^\mathrm{q} = 4.64\times 10^4$,
respectively. In the model where the $\alpha$-effect peaks at poles
($\alpha \sim \cos\theta$), the critical dynamo numbers are closer
to each other, $D_\mathrm{c}^\mathrm{d} = 2.36\times 10^4$ and
$D_\mathrm{c}^\mathrm{q} = 2.49\times 10^4$. Nevertheless, this
model also shows the preferred excitation of the
equator-antisymmetric magnetic fields.

Though the meridional flow is important for the latitudinal drift of
the magnetic fields in our model, the model does not belong to the
so-called advection-dominated dynamos. Figure \ref{f10} does not
show high concentration of surface fields towards the poles typical
of advection-dominated dynamo models. This is because of relatively
large magnetic diffusivity near the surface. The cycle period in our
model is not controlled by the meridional flow. A sufficiently long
cycle is realized mainly due to the low magnetic diffusivity in the
thin layer near the bottom boundary and partly due to the
diamagnetic pumping to this layer.

To probe for the effect of diamagnetic pumping, the computations
were repeated with the pumping switched off (for the model with
meridional flow and mid-latitude peaking $\alpha$-effect). The cycle
period at the critical dynamo number reduced to $P_\mathrm{cyc} =
0.47$, {\it i.e.}, a bit less than twice. Other model parameters are
influenced much more. Without diamagnetic pumping, the amount of
toroidal magnetic flux in the near-bottom region, producing the
$\alpha$-effect, decreases strongly. Accordingly, the critical
dynamo numbers increase by almost two orders of magnitude, to
$D^\mathrm{d}_\mathrm{c} = 2.47\times 10^6$ and
$D^\mathrm{q}_\mathrm{c} = 2.49\times 10^6$ with no clear preference
for dipolar parity. Without the pumping, the poloidal field at all
depths is of the same order as at the surface. Accordingly, the
ratio of amplitudes of the toroidal field to the surface polar field
decreased to about 43 from 920 in the model with diamagnetic
pumping.
\section{Summary}
Joint application of  diamagnetic pumping and a nonlocal
$\alpha$-effect strongly influences the solar-type dynamo model to
bring it generally closer to observations. The pumping concentrates
magnetic fields near the base of the convection zone where
diffusivity is relatively small. As a result, the solar cycle period
can be reproduced with the standard mixing-length value of eddy
diffusivity, $\eta_{_\mathrm{T}} \approx 10^{13}$\,cm$^2$s$^{-1}$,
in the bulk of the convection zone. The near-base concentration of
the poloidal field makes it possible for differential rotation to
wind a strong toroidal field over a solar cycle so that the toroidal
field amplitude in the dynamo model is about a thousand times
stronger than the surface polar field. The model produces
equator-antisymmetric global fields. The combination of diamagnetic
pumping and nonlocal $\alpha$-effect also resolves the theoretical
problem of the so-called catastrophic quenching of the alpha-effect
\cite{KO11b}.

The model without meridional flow does not, however, reproduce the
butterfly diagram of sunspot activity. The time-latitude diagram of
the toroidal field shows too slow equatorial drift and too broad
latitudinal distribution in this case (Figure~\ref{f8}). This can be
corrected by allowance for the meridional circulation {\em and} the
change of the latitudinal distribution of the $\alpha$-effect from a
pole-peaked to a mid-latitude-peaked profile (Figure~\ref{f10}).

The prescription for the $\alpha$-effect remains the main
uncertainty of the model. Operation of the nonlocal $\alpha$-effect
of the Babcock-Leighton type on the Sun is plausible from both
observational \cite{Dea10} and theoretical \cite{CMS95} evidences.
The mathematical \\ fo\-rmulation of this effect remains, however,
uncertain. If our model is considered as an inverse problem, it
predicts that the $\alpha$-effect is produced by deep-seated
toroidal fields and it is largest at mid latitudes.
 \begin{acks}
The authors are thankful to an anonymous referee for constructive
comments. This work was supported by the Russian Foundation for
Basic Research (projects 10-02-00148, 10-02-00391).
 \end{acks}

\end{article}
\end{document}